# Structural properties of plastically deformed $SrTiO_3$ and $KTaO_3$

Issam Khayr[1], Sajna Hameed[1,2], Jakov Budić[3], Xing He[1], Richard Spieker[1], Ana Najev[3], Zinan Zhao[4], Li Yue[4], Matthew Krogstad[6], Feng Ye[7], Yaohua Liu[7,8], Raymond Osborn[9], Stephan Rosenkranz[9], Yuan Li[4], Damjan Pelc[1,3], and Martin Greven[1]

[1]School of Physics and Astronomy, University of Minnesota, Minneapolis, MN, USA

[2]Max Planck Institute for Solid State Research, Stuttgart, Germany

[3]Department of Physics, Faculty of Science, University of Zagreb, Croatia

[4]International Center for Quantum Materials, School of Physics, Peking University, Beijing, China

[5]Beijing Academy of Quantum Information Sciences, Beijing 100913, China

[6]Advanced Photon Source, Argonne National Laboratory, Lemont, IL, USA

[7]Neutron Scattering Division, Oak Ridge National Laboratory, Oak Ridge, TN, USA

[8]Second Targe Station, Oak Ridge National Laboratory, Oak Ridge, TN, USA

[9]Materials Science Division, Argonne National Laboratory, Lemont, IL, USA

## ABSTRACT

Dislocation engineering has the potential to open new avenues toward the exploration and modification of the properties of quantum materials. Strontium titanate ($SrTiO_3$, STO) and potassium tantalate ($KTaO_3$, KTO) are incipient ferroelectrics that show metallization and superconductivity at extremely low charge-carrier concentrations and have been the subject of resurgent interest. These materials also exhibit remarkable ambient-temperature ductility, and thus represent exceptional platforms for studies of the effects of deformation-induced dislocation structures on electronic properties. Recent work on plastically deformed STO revealed an enhancement of the superconducting transition temperature and the emergence of local ferroelectricity and magnetism near self-organized dislocation walls. Here we present a comprehensive structural analysis of plastically deformed STO and KTO, employing specially designed strain cells, diffuse neutron and x-ray scattering, Raman scattering, and nuclear magnetic resonance (NMR). Diffuse scattering and NMR provide insight into the dislocation configurations and densities, as well as their dependence on strain. As in the prior work on STO, Raman scattering reveals evidence for local ferroelectric order near dislocation walls in plastically deformed KTO. Our findings provide valuable information about the self-organized defect structures in both materials, and they position KTO as a second model system to explore the associated emergent physics.

## I. INTRODUCTION

$SrTiO_3$ (STO) is one of the most extensively studied complex oxides and has been used as a model system to understand structural phase transitions [1-4], quantum criticality [5-7], and unconventional superconductivity [8]. After well over half a century of scientific efforts [9], there remain open questions about a range of quantum phenomena that arise in STO, including the emergence of superconductivity in the extreme dilute limit [10-13], the charge transport scattering mechanism that gives rise to $T^2$ resistivity [14-16], the interplay between superconductivity and ferroelectricity near a putative quantum critical point [5-7], and many more [17]. The closely related material $KTaO_3$ (KTO) shares important features with STO, such as a cubic perovskite structure at room temperature, a similar band gap, and incipient ferroelectricity at low temperatures. Moreover, two-dimensional superconductivity was recently discovered in KTO through electrolyte gating and at interfaces formed with EuO [18,19], which has made this material the subject of considerable renewed interest.

The properties of both STO and KTO are highly sensitive to strain, and mechanical deformation is thus an important experimental tuning parameter. In contrast to elastic deformation, in which stress and strain are linearly proportional (Hooke's law), much larger local strains can be obtained with plastic deformation, an irreversible process during which the atomic structure in a fraction of the crystal volume is severely altered. Although mostly used in the context of materials engineering, the possibility of plastic deformation in less ductile materials, such as ceramics, has been explored as well [20,21]. In fact, the field of dislocation mechanics has been extensively developed for over 60 years [22], but studies of the effects of dislocations on electronic properties of quantum materials are in their infancy. Comprehensive structural characterization is essential to gain a thorough understanding of possible long-range dislocation structures and, ultimately, of their effects on functional electronic properties. While transmission electron microscopy (TEM) [23,24] and x-ray diffraction [25,26] are commonly employed methods for the determination of dislocation densities in materials, techniques such as nuclear magnetic resonance (NMR) and resonant ultrasound spectroscopy have been utilized as well [27-30]. At sufficiently high deformation levels, dislocations intrinsic to the unstrained system can bunch or self-organize into novel structures that act as boundaries between relatively unaltered, dislocation-free regions [31].

It has been known for some time that STO is highly ductile at room temperature [32], an unusual property for a ceramic material. Electron microscopy experiments revealed that dislocations in the plastically-deformed material are highly correlated and self-organize into quasi-two-dimensional structures (dislocation walls) [32,33]. A recent study involving diffuse neutron and x-ray scattering, Raman scattering, and charge-transport measurements uncovered that superconducting and ferroelectric fluctuations in STO are enhanced upon plastic deformation [34]. This work furthermore linked the dramatic change in electronic properties to the formation of dislocation walls that separate essentially strain-free tilted domains, revealed as elongated Bragg reflections (asterisms) in neutron and x-ray scattering [34]. However, the evolution of these structures with strain has not been investigated. Furthermore, KTO has not been previously reported to be ductile at room temperature. As we show here, in agreement with a recent report [35], this perovskite can also be severely deformed in compression without fracture, joining STO and $KNbO_3$ as another perovskite ceramic with high room-temperature ductility [34,36]. In the present work, we perform a detailed comparative study of the structural properties of plastically deformed STO and KTO, with the principal aim to characterize the intricate dislocation structures that form during deformation and play a pivotal role in the modification of electronic properties.

We characterize the dislocation densities and their long-range correlations *via* neutron and x-ray diffuse scattering and complement the scattering data with a model calculation to obtain the strain-dependent change in linear dislocation density within the dislocation walls. Furthermore, using nuclear magnetic resonance (NMR), a complementary local probe, we estimate the fraction of the volume of a deformed KTO crystal that is significantly affected by the dislocation structures. Additionally, we present findings from Raman scattering measurements on plastically deformed KTO that indicate the emergence of local ferroelectric order, similar to the previous observations for plastically deformed STO [34]. Our comprehensive study sheds light on the universal structural effects of plastic deformation in these two perovskite oxides and provides a basis for further studies of their electronic properties.

This manuscript is organized as follows: Section II describes the sample preparation and experimental techniques. In Section III, we present our results: Stress-strain diagrams, diffuse neutron and synchrotron x-ray scattering data and modeling, as well as Raman scattering and NMR data. We then discuss and summarize our findings in Section IV.

## II. EXPERIMENTAL DETAILS

Undoped and 0.2% Nb-doped (carrier concentration ~$3 \cdot 10^{19}$ $cm^{-3}$) STO (lattice constant $a$ = 3.905 Å) as well as undoped KTO (lattice constant of 3.99 Å) crystals with dimensions of about 10 × 10 × 0.5 $mm^3$ were commercially purchased from MTI Corp. The single crystals were then cut and polished to suitable dimensions of about 1 x 1.5 x 0.5 $mm^3$ (mass ~2.5 mg) for the synchrotron diffuse x-ray and Raman scattering experiments. Separate, pre-aligned single crystals with dimensions of about 10 x 5 x 1 $mm^3$ (mass ~250 mg) were purchased from MTI Corp. and prepared for the diffuse neutron scattering and NMR experiments. In order to increase the likelihood of successful deformation, the small single crystals were cut into a rectangular shape with six facets polished to high precision. Initially undoped STO was oxygen-vacancy doped (OVD) by annealing samples in high vacuum at 800°C for 2 hours in the presence of a titanium getter, similar to procedures used in Refs. 9 and 34. The charge carrier concentration of the OVD samples was estimated from Hall number measurements to be approximately $10^{18}$ $cm^{-3}$. A total of 13 deformed STO and KTO samples were measured (see the sample information in Table I). Ten of these crystals were investigated *via* diffuse x-ray scattering: one undoped (STO_A), four oxygen-vacancy doped (OVD-STO_A, OVD-STO_B, OVD-STO_C, OVD-STO_D), and two 0.2% Nb-doped (NB-STO_A, NB-STO_B) STO samples, as well as three undoped (KTO_A, KTO_B, KTO_C) KTO samples. The Raman scattering measurements were performed on two of these deformed KTO samples (KTO_A, KTO_C) along with one undeformed sample (KTO_D). One OVD STO sample (OVD-STO_E) and one undoped KTO sample (KTO_E) were measured *via* diffuse neutron scattering. An additional undoped KTO sample (KTO_F) was used for NMR.

Three different strain cells were employed, all with the same operating principle: gas pressure on a piston within a cylinder is used to apply force to the sample in a controlled manner [34,37]. Two cells were used for *ex situ* deformation of samples which were subsequently measured at synchrotron and spallation neutron sources. A small cell with a maximum force of 350 N was used to deform nine STO and KTO samples for the x-ray and Raman scattering experiments. A large cell with maximum force of 5000 N was used to deform three STO and KTO samples for the NMR and neutron scattering measurements. For both cells, the sample deformation was determined with a linear variable transformer. A third cell, compatible with the goniometer setup at beamline 6-ID-D of the Advanced Photon Source, Argonne National Laboratory, was used for x-ray diffuse scattering measurements of the *in situ* deformed crystal STO_A. This cell enables measurements during crystal rotation (see below) and generates a maximum force of 150 N. In this measurement, the stress was determined directly from the gas pressure and sample geometry, and the strain was calculated from measurements of the sample dimension in the compression direction before and after deformation. All samples were deformed along the [0 1 0] crystallographic direction.

The diffuse x-ray scattering measurements were performed at beamline 6-ID-D of the Advanced Photon Source, Argonne National Laboratory. We used an 87 keV incident photon beam and a Pilatus 2M CdTe detector and utilized the standard crystal rotation method to collect data for many Brillouin zones in all momentum-space directions (see the Supplementary Information in Ref. 38). We note that only a fraction of the full rotation range was used for experiments with the strain cell in order to avoid scattering from cell components. The diffuse neutron scattering measurements were performed on the CORELLI spectrometer at the Spallation Neutron Source at Oak Ridge National Laboratory which uses a distribution of incident neutron energies ranging from 10-200 meV for wavevector magnitudes ranging from $Q$ = 2 to 8 r.l.u. No co-alignment or co-mounting was necessary due to the high efficiency of the spectrometer. The instrument has the possibility to separate elastic scattering, with a full-width-half-maximum (FWHM) energy resolution of 0.4-2.5 meV depending on the wavevector, from energy-integrated (to about 10 meV) "total" scattering [39]. $^{181}$Ta NMR measurements were performed at the University of Zagreb in a 12-T high-homogeneity magnet (Oxford Instruments), using Tecmag spectrometers and Tomco power amplifiers, and employed conventional spin-echo pulse sequences. The Raman scattering measurements were performed at Peking University in confocal backscattering geometry with a Horiba Jobin Yvon LabRAM HR Evolution spectrometer. BragGrate notch filters were used to acquire data down to around 10 cm$^{-1}$. The He-Ne laser line wavelength was 632.8 nm and the focused laser beam spot was 10 $\mu$m in diameter.

## III. RESULTS

We first present diffuse x-ray and neutron scattering results that unveil similarities in the self-organization of dislocations throughout the bulk of plastically deformed STO and KTO. The observed strain-dependent asterisms are modeled to shed light on the formation and evolution of dislocation walls across a wide range of strain levels. NMR spectroscopy is used to compare the bulk dislocation density of KTO before and after deformation on an absolute scale. Finally, Raman spectroscopy results reveal the emergence of ferroelectric fluctuations in KTO after deformation, complementary to previous findings for STO [34].

### A. Mechanical properties of STO and KTO

Figure 1 shows examples of room-temperature engineering stress-strain curves for STO and KTO. Additional stress-strain curves for Nb-STO and KTO are provided in the Supplemental Material [42]. Both materials are ductile at room temperature, which indicates that dislocations are present prior to deformation, and that their mobility is high [40]. A comparison of yield strengths with previous work reveals some variability. This and other sample information is compiled in Table 1 [32-34,41].We find no clear doping dependence of the yield strength for STO, consistent with findings in Ref. 34.

### B. Diffuse x-ray scattering

We used a custom-built pneumatic strain cell for an *in situ* diffuse x-ray scattering study of the effects of plastic deformation in STO (sample STO_A), with stress applied along [0 1 0]. We directly observe that Bragg reflections elongate azimuthally in the *HK*0 plane with increasing applied stress and time (on the scale of seconds to minutes). The duration of the applied stress was used here as a control parameter to achieve a range of deformation levels. Once the material reaches its yield strength, it can creep, *i.e.*, deform without further increasing the stress, before it starts to work-harden, at which point the stress needs to increase for the deformation process to continue. Therefore, instead of directly increasing the stress, we allowed the sample to be deformed at the same stress as a function of time. A supplemental movie of the response in the *HK*0 plane at different times and stresses is provided in [42]. As shown in Fig. 2(a,d,g), this is observed over a wide **Q** range and indicates the formation of low-angle tilt boundaries separated by dislocation walls. The newly formed diffraction pattern is referred to as an asterism (in analogy to the use of this term in astronomy to describe certain star patterns [43]). For example, asterisms have been observed in plastically deformed metals due to low-angle tilt domains separated by dislocation structures [44-46] and in STO [34].

Additionally, as shown in Fig. 2 for the (0 -2 2) and (-1 -2 2) reflections of STO, weak diffuse features along [1 1 0] and equivalent directions are seen to emerge with increasing stress, which indicates that the <1 1 0> slip planes are activated during plastic deformation. As shown in Ref. 34 and in the present work, the periodicity of these diffuse streaks is associated with the magnitude of the Burgers vector and the real-space distance between neighboring dislocations within deformation-induced dislocation walls. Such features are discussed in more detail in the next Section.

As seen in Fig. 3 for both OVD-STO and undoped KTO, the width of the asterisms at the (2 0 0) Bragg position tends to increase with increasing strain. For strains above about 2%, a splitting of Bragg reflections can be discerned, which indicates the formation of tilted domains separated by dislocation walls. As the strain further increases, the peak separation is seen to increase.

To the best of our knowledge, the structural effects of oxygen vacancy doping are negligible for the small vacancy concentrations that we used (below 1/1000). In general, one expects diffuse features to appear due to disorder. Given the small vacancy concentrations, this disorder can manifest as a diffuse cloud forming around Bragg reflections [47], but we observe no prominent difference in Bragg peak widths between undoped and doped STO. This is supported by diffuse scattering data (see Fig. S6 in [42]) for crystals with different doping levels, which do not show any discernible structural differences. Moreover, we do not find significant differences in ductility between undoped and doped samples.

### C. Diffuse neutron scattering

In order to isolate the static and dynamic components of the structural response, we employed the energy discrimination of the CORELLI instrument at the Spallation Neutron Source. We measured an OVD STO sample and a KTO sample. These data are complementary to the x-ray results, *e.g.*, with regard to the strain dependence of asterism widths. Additionally, given that both samples are deformed to 4.2%, this enables a direct comparison of the observed diffuse, cross-like features to quantify differences in the Burgers vector for the two materials.

Consistent with the x-ray results (Figs. 2 and 3), we observe an azimuthal broadening of Bragg reflections in the *HK*0 plane in the quasistatic diffuse neutron scattering data (Fig. 4). In both materials, the asterisms are confined to this plane, thus two of the four possible slip planes are activated. Although both crystals are deformed by 4.2%, the asterisms stretch over a larger $\boldsymbol{Q}$ range in STO, indicative if a higher dislocation density than in plastically deformed KTO. Another distinction between the two materials is in the diffuse features seen away from the Bragg reflections. In both cases, we observe streaks along [1 1 0], corresponding to the activation of the <1 1 0> slip planes. However, in the KTO sample, streaks are observed along [1 -1 0] as well. The diffuse neutron scattering data for plastically deformed OVD STO in Ref. 34 (also deformed by 4.2%) also show streaks in both directions, which suggests that the observed variability regarding activated slip planes is sample-dependent and not compound specific.

In the total scattering channel, diffuse features indicative of deviations from average structure can be masked by phonon scattering, especially when the streaks do not extend far away from the Bragg reflections. These effects are more severe in the case of KTO, which exhibits a strongly anisotropic and relatively soft transverse acoustic phonon branch [48]. We note that the diffuse streaks in the total scattering channel are still observed, but that they are partially masked by radial streaks due to instrument artifacts (see Fig. S3 [42]).

**D. Modeling of fine structure**

We now employ a simple model to describe the effects of stress observed in the x-ray and neutron scattering data. As shown in Fig. 3, we observe an overall increase in the width of asterisms as the deformation level increases. In order to model this behavior, we calculate the kinetic scattering from a dislocation wall, following Ref. [34]. For a single dislocation, the displacement field at a position $\mathbf{z} = x + \mathrm{i}y$ in a complex plane perpendicular to the dislocation direction is [40]

$$\mathbf{u}(\mathbf{b},\mathbf{z}) = \frac{-\mathrm{i}\mathbf{b}}{2\pi}\log\mathbf{z} + \frac{1}{4(1-\sigma)}\left(\frac{\mathrm{i}\mathbf{b}}{2\pi}\log\mathbf{z}\bar{\mathbf{z}} - \frac{\mathrm{i}\bar{\mathbf{b}}}{2\pi}\frac{\mathbf{z}}{\bar{\mathbf{z}}}\right), \tag{1}$$

where $\mathbf{b}$ is the Burgers vector, σ is the Poisson ratio, and the overline refers to the complex conjugate of the variable. In our calculation, we revise the above expression to orient the dislocation along [1 1 0] through a rotational transformation. The displacement field for a wall can then be easily calculated by summing the displacements of an array of dislocations in the (1 1 0) plane, and we determine the kinetic scattering by simple lattice summation. A 400 x 400 square lattice is used in this model.

The parameters that are adjusted to replicate the asterisms observed in the experiment are the real-space distance between dislocations and the magnitude of the Burgers vector, which will determine the linear dislocation density. Figure 5(a)-(d) displays the calculated two-dimensional contour plots, centered at the (2 0 0) Bragg position, where the distance *d* between dislocations is varied and the Burgers vector magnitude is fixed to $2\sqrt{2}a$, where $a$ is the lattice constant. The corresponding result for the (4 0 0) Bragg position is provided Fig. S4 [42]. Two effects can be observed upon changing the distance between dislocations, which increases the dislocation density within a dislocation wall: (*i*) an increase in the azimuthal spread of the asterisms and (*ii*) a change in the number of streaks per reciprocal lattice unit, or periodicity, of the fine structure (cross-like features), which can be seen in the one-dimensional cuts along [*H K* 0] in Fig. 5(e-h). The linecut at $H = 0$ conveys observation (i): the spread of the asterisms increases with increasing dislocation density. The linecut at $H = 1.875$ r.l.u. conveys observation (ii): the momentum space distance between the cross-like features increases with increasing dislocation density and fixed Burgers vector. In the most extreme case, the streaks are resolution-limited, with no apparent features for $d = 151$, but as $d$ is reduced, more peaks are discerned in the linecuts due to scattering from the dislocation walls. The real-space schematic in Fig. 5 attempts to give a physical picture, namely that the increased spread of the asterisms implies larger tilt angles between domains separated by dislocation walls as the Burgers vector remains constant.

In order to study the structure of the diffuse streaks without modifying the angular spread between the main Bragg peaks, we simultaneously changed $d$ and the magnitude of the Burgers vector $b$ to keep the linear dislocation density constant within the wall. As seen from Fig. 6(e), this significantly modifies the periodicity of the diffuse features around the (2 0 0) Bragg position (for the (4 0 0) reflection, this is shown in Fig. S5 [42]). Additionally, as $d$ is decreased, the diffuse features extend further away from the Bragg reflections, since the strain field changes over shorter real space distances.

In Fig. 7, we explore this behavior for KTO. For the 4.2% deformed sample studied *via* neutron scattering, the angular spread of the (2 0 0) Bragg reflection is 3.4 degrees (see Figs. 4 and 8). This value only provides the linear dislocation density and cannot be used to determine the size of the Burgers vector magnitude without additional information. We therefore use a comparison between the diffuse streaks and model calculation to determine the Burgers vector: we keep the dislocation density constant and simultaneously adjust $d$ and $b$ to see which periodicity best matches the data. From Fig. 4(c), we discern about four to five streaks over the reciprocal spacing between the split (200) Bragg reflection, which is similar to the density of streaks in Fig. 7(c) and suggests that the Burgers vector for KTO is $5\sqrt{2}a$.

We now summarize the x-ray and neutron scattering datasets for undoped, OVD, and Nb-doped STO, as well as undoped KTO, in order to determine how the asterisms change with the strain level, and whether mobile carriers substantially affect the ductility or dislocation self-organization in the case of STO. To this end, we use the angular spread of the asterisms (similar to information extracted from rocking curve measurements [25,26]) as a figure of merit to directly relate strain to dislocation density. In Fig. 8, each symbol represents the FWHM of the angular spread averaged over ten or more Brillouin zones for each sample. The raw data for each deformed sample are provided in [42]. Since the asterisms consist of two principal peaks (with the exception of sample OVD-STO_A shown in Fig. S6a, which was deformed to only 0.6%), data such as the one-dimensional cuts in Fig. 3 were fit to two gaussians, and the FWHM was defined as

$$\frac{1}{2}\,(f_1 + f_2) + (c_2 - c_1), \tag{2}$$

where $f_{1,2}$ are the FWHM values of the individual peaks and $c_{1,2}$ are the peak centers. The error bars (mostly within the size of the symbols in Fig. 8) do not represent fit errors, but rather standard deviations due to the averaging over multiple Brillouin zones. Overall, there is very little deviation up to high momentum transfers (as high as $Q = 8$ r.l.u.), although there might be some bias, since the reflections were chosen based on how well the two-peak fit could be performed. Some samples show little variation from zone to zone (as is the case of OVD-STO_C), while others show more variation (KTO_C) in peak width. In addition to extracting the FWHM of the asterisms from experiment, we perform the corresponding analysis of the calculated diffuse scattering. Here, the Burgers vector magnitude is fixed to $2\sqrt{2}a$, while the distance between dislocations is varied within the dislocation wall. This effectively varies the linear dislocation density, which is scaled by $a$. These results are included in Fig. 8, with the modeled asterism width plotted as a function of linear dislocation density rather than as a function of strain. As a reference, we note that setting the distance between dislocations to 16 unit cells corresponds to ~ 4.2% deformation, in agreement with Ref. 34. Although KTO has a different Burgers vector magnitude, it is still included in this plot to convey the similar qualitative (linear) relationship between dislocation density and strain.

**E. Raman scattering measurements of KTO**

The parallels between our results for STO and KTO extend beyond the observed mechanical and structural similarities evident from stress-strain curves and diffuse scattering. Our Raman scattering results for plastically deformed KTO reveal further similarities with prior work on STO [34] regarding the emergence of phonon modes that indicate the presence of broken inversion symmetry in a non-zero volume fraction of the bulk.

As shown in Fig. 9, three undoped KTO samples were measured. The undeformed sample shows all the expected phonon modes that are allowed considering the high symmetry of the material, in good agreement with prior Raman results [48-51]. Previous inelastic neutron scattering work investigated the transverse-optic $TO_1$ mode, which softens, but does not condense to zero at low temperatures, signifying the material's proximity to a ferroelectric phase [5-7]. Due to the high symmetry of KTO and the nature of Raman scattering, the various TO modes are not observable in pristine, unstrained samples. However, this probe can be extremely sensitive to inversion-symmetry breaking, even in a small volume fraction of a sample, which is precisely what we observe after plastic deformation to 2.8% and 7.0%. Upon cooling, the hard $TO_2$ and $TO_4$ modes (~200 $cm^{-1}$ and ~545 $cm^{-1}$) are seen to emerge below about 50 K. Additionally, we observe a deformation-induced low-energy soft $TO_1$ mode (below 50 $cm^{-1}$) with a spectral-weight buildup at low temperatures, indicative of the emergence of ferroelectric fluctuations near dislocation walls. These findings are similar to prior results for STO [34].

**F. NMR measurements of KTO**

Diffuse scattering and Raman spectroscopy provide detailed insight into the makeup of the dislocation structures, but it is difficult to estimate the volume fraction that is significantly affected by the local strain. In order to obtain this important information, we use NMR as a complementary local bulk probe. In particular, the measurements of the $^{181}$Ta NMR spectra of KTO before and after deformation shown in Fig. 10 provide a simple way to estimate the strained volume fraction, and our calculation of the spectra based on the strain fields of dislocation walls is in quantitative agreement with experiment.

The basic nuclear spin Hamiltonian consists of the Zeeman and the quadrupolar terms, representing the effect of an applied field $\boldsymbol{H}$ and the interaction of the nuclear quadrupole moment with the electrical field gradient (EFG) tensor, respectively:

$$\mathrm{H}_{\mathrm{spec}} = -\gamma\hbar\mathbf{I}\cdot\left(\hat{\mathbf{1}}+\hat{\mathbf{K}}\right)\cdot\mathbf{H} + \frac{\mathrm{e}^2 qQ}{4I(2I-1)}\left(3I_z^2 - I(I+1) + \frac{\eta}{2}\left(I_+^2 + I_-^2\right)\right) \tag{3}$$

Here, the notation is conventional: **I** is the nuclear spin, $\hat{\mathbf{K}}$ is the NMR shift tensor and $\hat{\mathbf{1}}$ is the unit tensor; $q$ is the principal component of the EFG tensor and η represents the EFG asymmetry parameter; γ and $Q$ are the nuclear gyromagnetic ratio and quadrupolar moment. In case of axial symmetry, η is zero and the eigenenergies can be simply calculated assuming that the quadrupolar term is a small perturbation. The NMR frequencies to first order in perturbation theory are

$$\nu_{NMR} = \nu_L \pm n\nu_Q(3\cos^2\theta - 1)/2, \tag{4}$$

where $\nu_L$ is the unperturbed Larmor frequency, $n = 0, 1, 2,$ or $3$ for a spin $7/2$ nucleus such as $^{181}$Ta, the quadrupole frequency $\nu_Q$ is given by $e^2 qQ/4I(2I - 1)$, and θ is the angle between the magnetic field and the principal axis of the EFG tensor. To first order, the frequency of the central line ($n = 0$) is unaffected by the quadrupolar term, and is therefore determined by the second-order perturbation term

$$\nu^{(2)}_{central} = \nu_L - \frac{15}{16}\frac{\nu_Q^2}{\nu_L}\sin^2\theta\,(9\cos^2\theta - 1). \tag{5}$$

In contrast, the satellite transitions ($n \neq 0$) show a first-order effect, which is thus easier to observe. Since KTO has a cubic structure, the EFG tensor at the Ta site is zero by symmetry in the absence of local strain, and the resonant frequencies for all $n$ are the same. $^{181}$Ta NMR is thus a sensitive probe of the volume fraction of unit cells that show local deviations from the cubic structure: only the unit cells where such deviations exist will show quadrupolar satellite lines. The measured spectrum of the plastically deformed sample clearly shows the presence of two components (Fig. 10): a narrow central line, and a broad signal that results from a superposition of satellite lines of nuclei with different $\nu_Q$. It is straightforward to extract the absolute volume fraction of unit cells affected by the dislocation-induced strain fields from the weighted ratio of the intensities of the central and satellite lines. For a spin-7/2 nucleus such as $^{181}$Ta, the relative intensities are

$$N_{central} = \frac{2}{21}, \qquad N_1 = \frac{\sqrt{15}}{42}, \qquad N_2 = \frac{\sqrt{12}}{42}, \qquad N_3 = \frac{\sqrt{7}}{42}.$$

The absolute volume fraction of the affected unit cells is then

$$W = \frac{I_{satellites}}{I_{central}}\frac{N_{central}}{N_1 + N_2 + N_3} \approx 5.0 \cdot \frac{I_{satellites}}{I_{central}}. \tag{6}$$

A straightforward integration of the measured intensities yields a volume fraction of 1.2%. Note that this is a lower limit: the satellite signal could still exist at frequencies beyond the integration range, but it becomes indistinguishable from the noise.

Similar to diffuse scattering, the NMR spectrum can be modeled using the strain fields around dislocation walls. For simplicity, we neglect the shear strain components, and assume that the quadrupolar frequency is related to the strain *via* [52]

$$\nu_Q = A\left(\varepsilon_a - \frac{1}{2}\varepsilon_b\right), \tag{7}$$

where $A$ is a coupling constant, $\varepsilon_a$ is the larger and $\varepsilon_b$ is the smaller of $\varepsilon_{xx}$ and $\varepsilon_{yy}$ at each site. The spectrum is then calculated numerically by summing over a 2D grid of unit cells that includes a dislocation wall, as follows. Let us consider a two-dimensional slice of a crystal with a dislocation wall in the $y$ direction at $x = 0$. We create a two-dimensional array of unit cell coordinates ranging from $-N_x$ to $N_x$ in the $x$ direction and $-N_y$ to $N_y$ in the $y$ direction, and then calculate the strain field and quadrupolar frequency at each point. The theoretical NMR signal intensity can then be obtained by summing the individual spectra of all unit cells within the array. Each individual spectrum is assigned a small effective gaussian width in order to minimize numerical noise in the summed spectrum. We include angles $\theta$ and $\theta + \pi/2$, as we assume that perpendicular dislocation walls are present in the crystal, as concluded from the diffuse scattering data. The calculated spectra are adjusted to fit the data using $A$, $N_y$ and $\theta$ as free parameters; $N_y$ gives the effective distance between dislocation walls and is thus closely related to the volume fraction. We obtain the best agreement with the data for $N_y = 32$, which then gives a measure of the average distance between dislocation walls. The lattice strains are therefore small in large regions of the sample, in agreement with the appearance of well-defined asterisms in diffraction.

## IV. DISCUSSION AND CONCLUSION

While STO and KTO display qualitatively similar mechanical properties, there exist subtle differences. We observe a work-hardening regime, characterized by an upturn in the stress-strain curve, in all STO samples that were deformed beyond ~2%. This occurs for sufficiently high applied stress above the yield point. As self-organized dislocation structures form, it is increasingly difficult for additional dislocations to move freely, which results in an increase of the stress required to further deform the material. In contrast, we find for all KTO samples that the stress-strain curve is approximately linear in the plastic regime, with no obvious upturns. There might be no work-hardening regime at all, but this seems unlikely given that dislocation multiplication and nucleation, expected to occur before the sample work-hardens, is more energy-costly compared to the motion and self-organization of already existing dislocations [22]. It is more probable that work hardening in KTO commences immediately above the yield stress and continues up to the failure stress, similar to conclusions arrived at in Ref. 35.

Discrepancies in the yield strength and elastic moduli for both STO and KTO can possibly arise from differences in measuring stress and strain of deformed samples and from imperfections in cutting and polishing samples, which can alter the effective area on which the force is applied and, hence, the effective stress. Given the high precision of our sample-dimension and applied-force measurements, the uncertainties in the individual yield strength estimates are about 1% or less. On the other hand, we find the sample-to-sample variation of the yield strength estimates to be significantly larger (~20-30%, see Table 1), likely due to small sample shape imperfections. Since we only determined the yield strengths for a subset of the studied samples, the overall systematic error is large. In the linear regime of the stress-strain diagrams, we find significant deviations from the expected elastic moduli for both STO and KTO of approximately an order of magnitude: *e.g.*, about 20 GPa for both STO and KTO (see Table 1), compared to the expected values of 272 GPa [53] and 238 GPa [54], respectively. These discrepancies are, again, most likely linked to our sample preparation process and will be further investigated in future studies. We also observe differences in the slopes of the stress-strain curves pre- and post-deformation. One would expect to find similar elastic moduli in both regimes [55], but it is possible that varying the gas pressure ramp rate affects the slopes. Typically, we used a faster ramp rate to release pressure, which could cause the slope to increase post-deformation.

Both materials exhibit remarkable room-temperature ductility, without much of an effect of doping on ductility or yield strength in the case of STO. In addition to tracking the sample displacement with a linear variable transformer (LVT), we also measured the sample dimensions before and after plastic deformation with micron resolution, which confirmed the dramatic deformation that these materials can undergo. These values typically agreed within 1% with those determined from the measured voltage change of the LVT coils. Finally, we note that we observe streaks on the surfaces of the deformed crystals that indicate the termination of dislocation walls (see also the supplemental material of Ref. 34). Although we observe some variability in the elastic moduli and yield strengths, we find good reproducibility of the ability to deform the materials without fracture, and we achieve a high level of control of deformation to a specific level due to the high resolution of the LVT coils and the reliable application of stress using gas pressure.

The diffuse scattering data reveal the emergence of asterisms in the *HK*0 plane after deformation, with two resulting primary peaks in each Brillouin zone, corresponding to the two types of domains that form and are separated by dislocation walls (Figs. 2-4). The relative intensity of the two Bragg peaks is sample-dependent and can differ by more than a factor of two at severe deformation levels. This domain imbalance can appear for multiple reasons, *e.g.*, slight deviations from the nominal [0 1 0] applied stress direction or minor imperfections of polished surfaces.

It is known that undeformed STO undergoes a cubic-to-tetragonal transition below about 100 K [1-4], which leads to the appearance of superstructure Bragg peaks at (½ ½ ½)-type positions. Similar to the integer Bragg reflections, these superstructure peaks show asterisms in deformed samples (see Fig. S3 of Ref. 34). The structural transition appears to be unaffected by plastic deformation, as the temperature dependence of the superstructure intensity does not show significant changes compared to the undeformed case. Undeformed KTO remains cubic at all temperatures, and we do not observe additional Bragg peaks that would suggest such symmetry-lowering in portions of deformed samples. We therefore rule out the possibility that tetragonal distortions emerge in the vicinity of the domain walls, or that macroscopic tetragonal domains are formed during deformation.

Upon comparing the x-ray and neutron scattering profiles (Figs. 3 and 4), we notice that the cross-like diffuse features corresponding to the activation of slip planes are clearly resolved with the latter probe (in both the quasistatic and total scattering channels), while they are barely discernible with the former, apart from the x-ray dataset for the *in situ* deformed STO crystal (Fig. 2). One possible reason for this is that the high sensitivity of neutrons to oxygen displacements plays a role in the observation of the diffuse streaks. However, since the streaks are due to relatively long-range strain fields around dislocation walls that involve the displacement of all atoms within a given unit cell, it is not clear whether this should be a factor. We believe that the main reason for the better visibility of diffuse streaks in the neutron scattering data are differences in sample geometry, which can dictate the number of slip planes that are activated. For a perfectly aligned and polished sample, the four [1 1 0]-type slip planes are nominally equivalent. However, in the case of small misalignments during the deformation process a set of only two slip planes might dominate, which would likely enhance the visibility of the diffuse streaks. The samples used in the neutron scattering measurements have typical dimensions of 10 x 5 x 1 $mm^3$, *i.e.*, there is an order of magnitude difference between the shortest and longest dimension, which might bias the formation of slip planes. In contrast, the dimensions of the x-ray samples are all similar, typically 1 x 1.5 x 0.5 $mm^3$. Yet other effects could also influence the slip plane activation, such as a tiny misalignment of the surfaces upon which the stress is applied (either due to sample polishing imperfections or misaligned anvils). To our knowledge, the structural features do not appreciably change over time in deformed samples, but in principle, it is possible that some relaxation occurs, which could make it harder to observe the diffuse features in *ex situ* deformed samples.

Based on visual comparison of the diffuse neutron scattering results in Fig. 4, we find a qualitative difference in the periodicity of the fine structure of STO and KTO, indicative of different Burgers vector magnitudes. For plastically deformed STO, the calculated diffuse scattering patterns for different Burgers vector magnitudes indicate that a distance of $16a$ between neighboring dislocations within a dislocation wall and a Burgers vector magnitude of $b_0 = 2\sqrt{2}a$ (Fig. 6b)) matches the observed fine-structure periodicity (Fig. 4a). In the case of plastically deformed KTO, the Burgers vector magnitude is $b_0 = 5\sqrt{2}a$, larger by a factor of 2.5 compared to STO (Fig. 7). This, in turn, implies that dislocation bunching is more pronounced in KTO, which will hinder dislocation mobility and thus require higher stresses for the material to deform. This observation is in qualitative agreement with the possibility of an extended work-hardening regime, as seen from the stress-strain curves for KTO.

From Fig. 8, we deduce that there is a universal linear relation between strain, the FWHM of asterisms, and the linear dislocation density within a dislocation wall. This is consistent with the expectation that higher strain levels lead to larger angular tilts between domains, which then results in larger angular spreads of the Bragg reflections. The dependence of the asterism width on strain was fit to a power-law form with exponent $p$. Within error, we find a linear relationship ($p = 1.0 \pm 0.2$). As shown in Fig. 8, if we combine this result with insights from the model calculation, we can additionally conclude that there exists a linear relationship between strain and dislocation density in the walls. We emphasize that there is some scatter in the results summarized in Fig. 8. For example, two 2.4% deformed STO samples (OVD-STO_B and NB-STO_A) vary in angular spread by 2.8 degrees. We do not ascribe this discrepancy to the difference in carrier concentration or doping type, but rather attribute it to systematic errors associated with the sample preparation process. For example, deviations of the stress direction from [0 1 0] can be introduced while cutting or polishing samples, and this can lead to a preferential arrangement of the tilted domains during deformation and affect the angular spread of the diffuse scattering.

Another point to note is that, although the Burgers vector magnitude for plastically deformed KTO is larger than for STO, the KTO results in Fig. 8 approximately follow the same linear relation. This indicates that, regardless of the magnitude of the Burgers vector, an increase in dislocation density within a dislocation wall is a direct result of an increase in the deformation level. This is a nontrivial conclusion: it would also be possible for the number of walls to increase with increasing stress, without substantial changes of the dislocation density within each wall. Yet once the walls are formed at relatively low deformation levels, they likely act as traps for mobile dislocations, which leads to the observed strain dependence. The fact that the dislocation density within the walls can be engineered with strain is of potential importance for applications that involve the tuning of electronic properties, such as superconductivity and ferroelectricity [34] as well as emergent magnetism [56].

From Raman scattering, we observe evidence for inversion-symmetry breaking in a nonzero volume fraction of KTO, consistent with prior observations for STO [34]. We are able to estimate this fraction from our NMR data and find it to be rather small (about 1 %), which indicates that most of the deformed crystal retains high symmetry. Our key NMR result is therefore that the overwhelming volume of deformed crystals is essentially unaffected by the dislocation structures, while the large strains are concentrated in a small effective volume. This is also in line with the scattering data: given that the Bragg peaks remain well-defined in the radial direction, a significant fraction of unit cells must retain their original structure. NMR, however, provides a straightforward quantitative estimate of the fraction of unit cells affected by deformation. Nevertheless, these dislocation structures have significant implications for the system in their vicinity, causing inversion-symmetry breaking, local ferroelectric and magnetic order, as well as enhanced superconductivity in deformed STO [34, 56]. In addition, the increase in spectral weight of the deformation-induced phonon modes with increasing deformation level seen in the Raman data indicates that it is possible to manipulate the strength of dislocation-related ferroelectric correlations.

In conclusion, we present a comprehensive characterization of the effects of plastic deformation on the classic perovskites STO and KTO that involves x-ray, neutron and Raman scattering, as well as NMR and modeling. The diffuse x-ray and neutron scattering data reveal universal dislocation mechanics and formation of dislocation walls in both compounds. Modeling of the scattering data confirms that dislocations are not randomly distributed throughout the bulk, but rather highly correlated, forming self-organized domain walls. Given the high ductility of these perovskite oxides already at room temperature, plastic deformation must be mediated *via* both dislocation multiplication and mobility, which are responsible for an initial increase in the dislocation density (multiplication) and the linear increase in dislocation density within the formed dislocation walls (mobility). Our Raman scattering data provide evidence for deformation-induced ferroelectric fluctuations in KTO, analogous to prior observations for STO [34]. Given the close proximity of ferroelectricity and superconductivity in STO, and the enhancement of superconductivity in plastically deformed samples, it will be interesting to search for evidence of superconductivity in doped, plastically deformed KTO. As of yet, there have only been reports of two-dimensional superconductivity in (undeformed) KTO, achieved *via* ionic gating [18] and interfacing with another crystalline material [19]. Our NMR measurements of KTO reveal that, although the crystal structure near the dislocation walls is dramatically altered, only a comparatively small volume fraction is highly strained. In combination with our diffuse scattering results, this implies that, as the strain level is increased, the unstrained volume of the crystal becomes *more* homogenous after deformation, given that dislocations migrate to the walls. Overall, our comprehensive approach highlights the vast potential of strain engineering in the context of strongly correlated materials.

## ACKNOWLEDGEMENTS

We thank C. Leighton, R. M. Fernandes, A. McLeod, B. Kalisky, and A. Klein for discussions and comments. This work was supported by the Department of Energy through the University of Minnesota Center for Quantum Materials, under grant number DE-SC-0016371, and by the Croatian Science Foundation, under grant number UIP-2020-02-9494. The work in Zagreb used equipment funded in part through project CeNIKS co-financed by the Croatian Government and the European Union through the European Regional Development Fund - Competitiveness and Cohesion Operational Programme (Grant No. KK.01.1.1.02.0013). Work at Argonne (RO and SR) was supported by the U.S. Department of Energy, Office of Science, Basic Energy Sciences, Materials Sciences and Engineering Division. This research used resources at the Spallation

Neutron Source, DOE Office of Science User Facilities operated by Oak Ridge National Laboratory, and at the Advanced Photon Source, a DOE Office of Science User Facilities operated by Argonne National Laboratory.

| Name | Doping | Strain % | Probe | Yield Strength | Elastic Modulus |
|---|---|---|---|---|---|
| **OVD-STO_A** | Oxygen vacancies | 0.6 | x-ray | - | - |
| **OVD-STO_B** | Oxygen vacancies | 2.4 | x-ray | - | - |
| **OVD-STO_C** | Oxygen vacancies | 4.1 | x-ray | - | - |
| **OVD-STO_D** | Oxygen vacancies | 5.8 | x-ray | 80 MPa | 20 GPa |
| **OVD-STO_E** | Oxygen vacancies | 4.2 | neutron | - | - |
| **NB-STO_A** | 0.2% Nb | 2.4 | x-ray | - | - |
| **NB-STO_B** | 0.2% Nb | 5.8 | x-ray | 100 MPa | 40 GPa |
| **STO_A** | Undoped | 2.0 | x-ray | - | - |
| **KTO_A** | Undoped | 2.8 | x-ray, Raman | - | - |
| **KTO_B** | Undoped | 4.4 | x-ray | - | - |
| **KTO_C** | Undoped | 7.0 | x-ray, Raman | 110 MPa | 20 GPa |
| **KTO_D** | Undoped | 0 | Raman | - | - |
| **KTO_E** | Undoped | 4.2 | neutron | 175 MPa | 29 GPa |
| **KTO_F** | Undoped | 2.9 | NMR | - | - |
| **STO (Ref. 32, 33)** | Undoped | - | - | 150 MPa | - |
| **NB-STO (Ref. 34)** | 0.2% Nb | 5.3 | - | 150 MPa | 270 GPa |
| **NB-STO (Ref. 34)** | 2% Nb | 8.7 | - | 230 MPa | 200 GPa |
| **STO (Ref. 41)** | Undoped | 19 | - | 45 MPa | 228 GPa |
| **KTO (Ref. 35)** | Undoped | 6.3 | - | 274 MPa | - |

Table I: Sample list with respective doping information, deformation levels, and measurements that were performed. The yield strength was only determined for four of the 13 samples measured in the present work, since the focus was on achieving distinct deformation levels. Elastic moduli for the four measured samples are calculated based on the stress strain curves in Figs. 1 and S1 in the low-strain elastic regime. The elastic moduli were not provided in Ref. 32, 33, and 35.

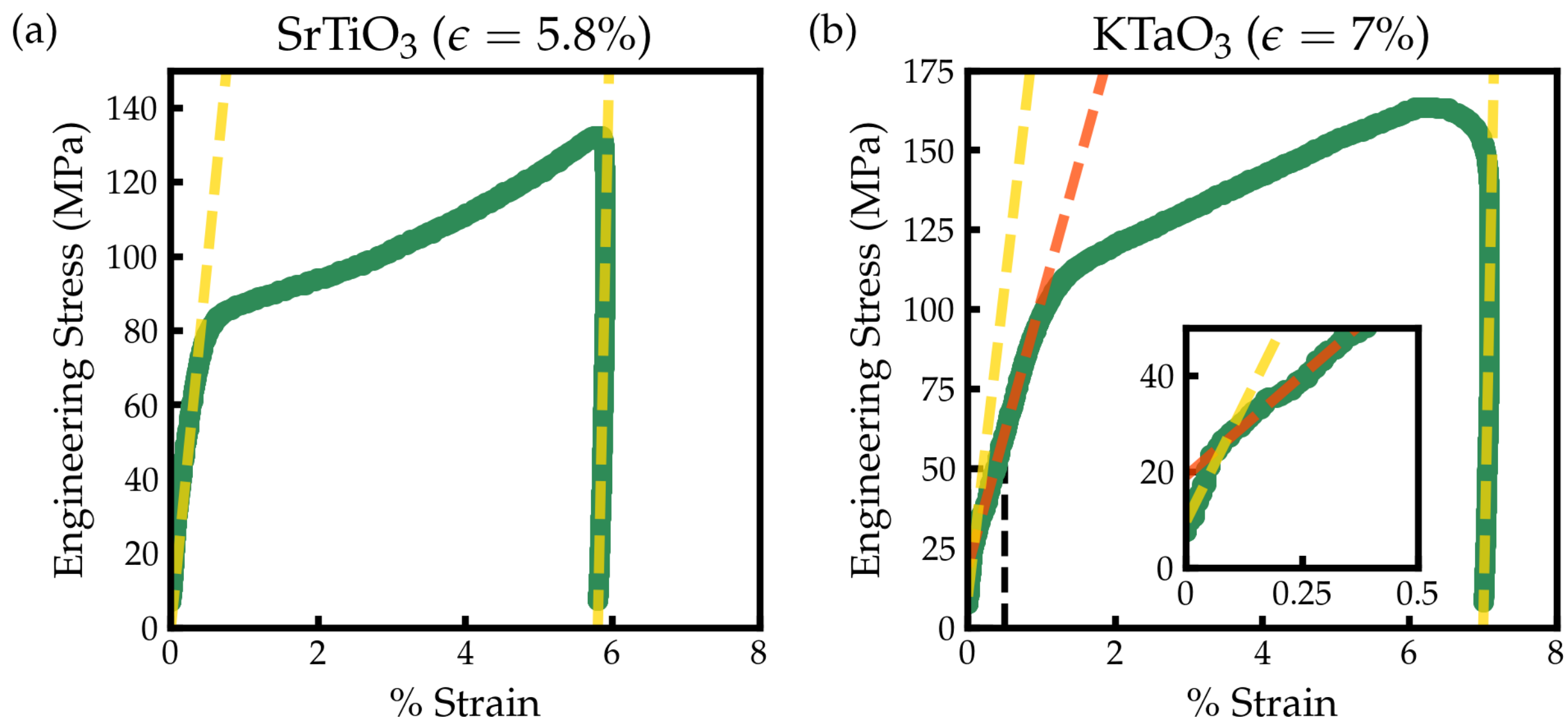

Fig. 1: Representative room-temperature stress-strain curves for samples (a) OVD-STO_D and (b) KTO_C, with compressive uniaxial stress along [0 1 0]. The respective inflection points around 80 MPa and 120 MPa signify the beginning of irreversible plastic deformation. The dashed lines indicate estimates of elastic moduli (see also Table S1). The inset to (b) shows that there appear to be two linear regimes, represented by the red and yellow dashed lines; this is discussed in the main text.

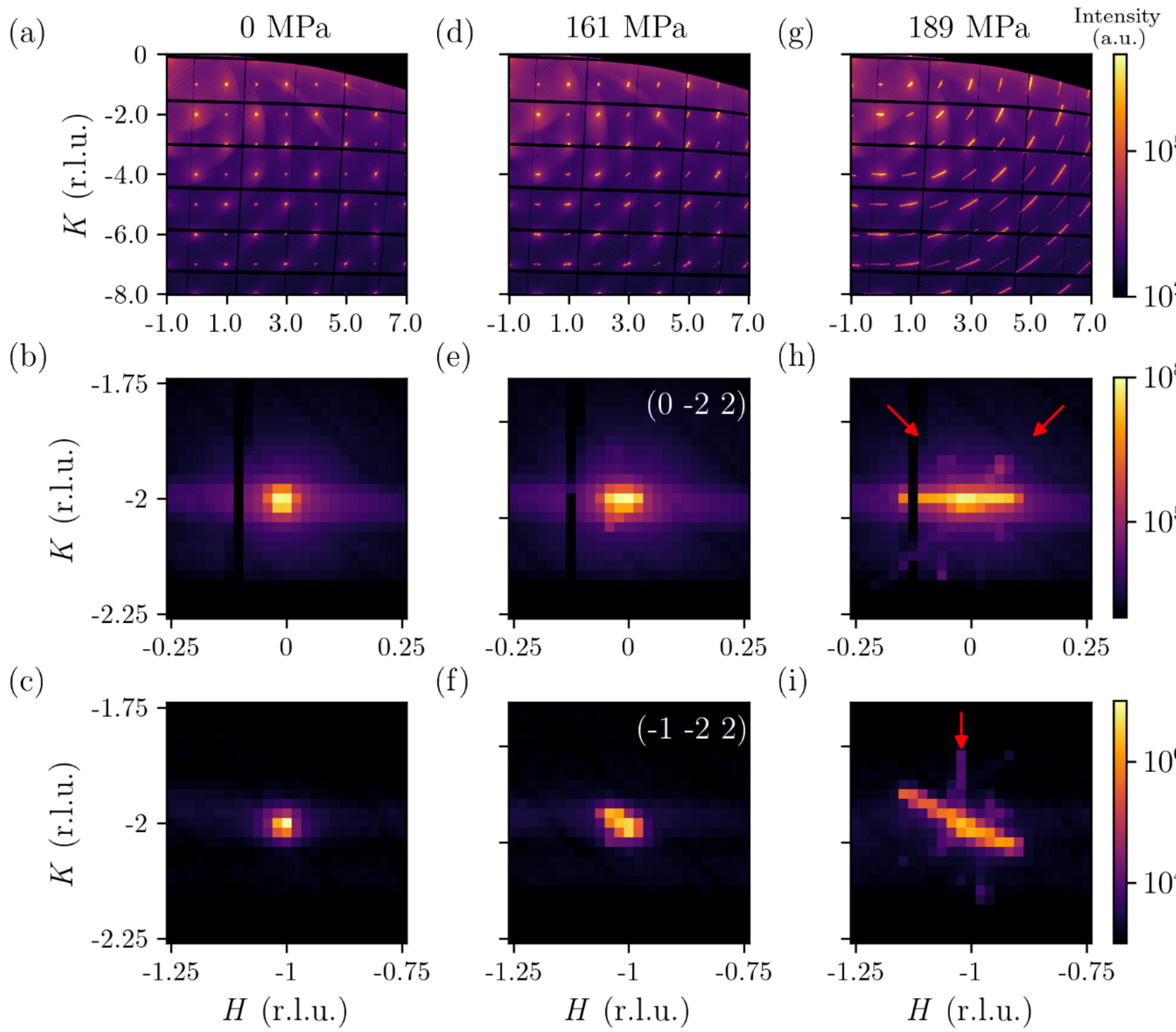

Fig. 2: Diffuse x-ray scattering data for sample STO_A, deformed *in situ* along [0 1 0] at room temperature, with stress levels: (a-c) 0 MPa, (d-f) 161 MPa, and (g-i) 189 MPa. Reflections in the *HK*0 plane start out as sharp Bragg peaks and gradually smear out as the sample is plastically deformed (first row). Diffuse cross-like features, indicated by the red arrows in (h) and (i), are observed at the (0 -2 2) and (-1 -2 2) reflections and indicate the activation of slip planes. Gaps in the data (black streaks) indicate gaps in the detector. For the maximum stress level of 189 MPa, the *HK*1.9 and *HK*2.1 planes are shown in Fig. S2 [42] to demonstrate that the weak features are confined to the *HK*2 plane and strictly oriented along [1 1 0].

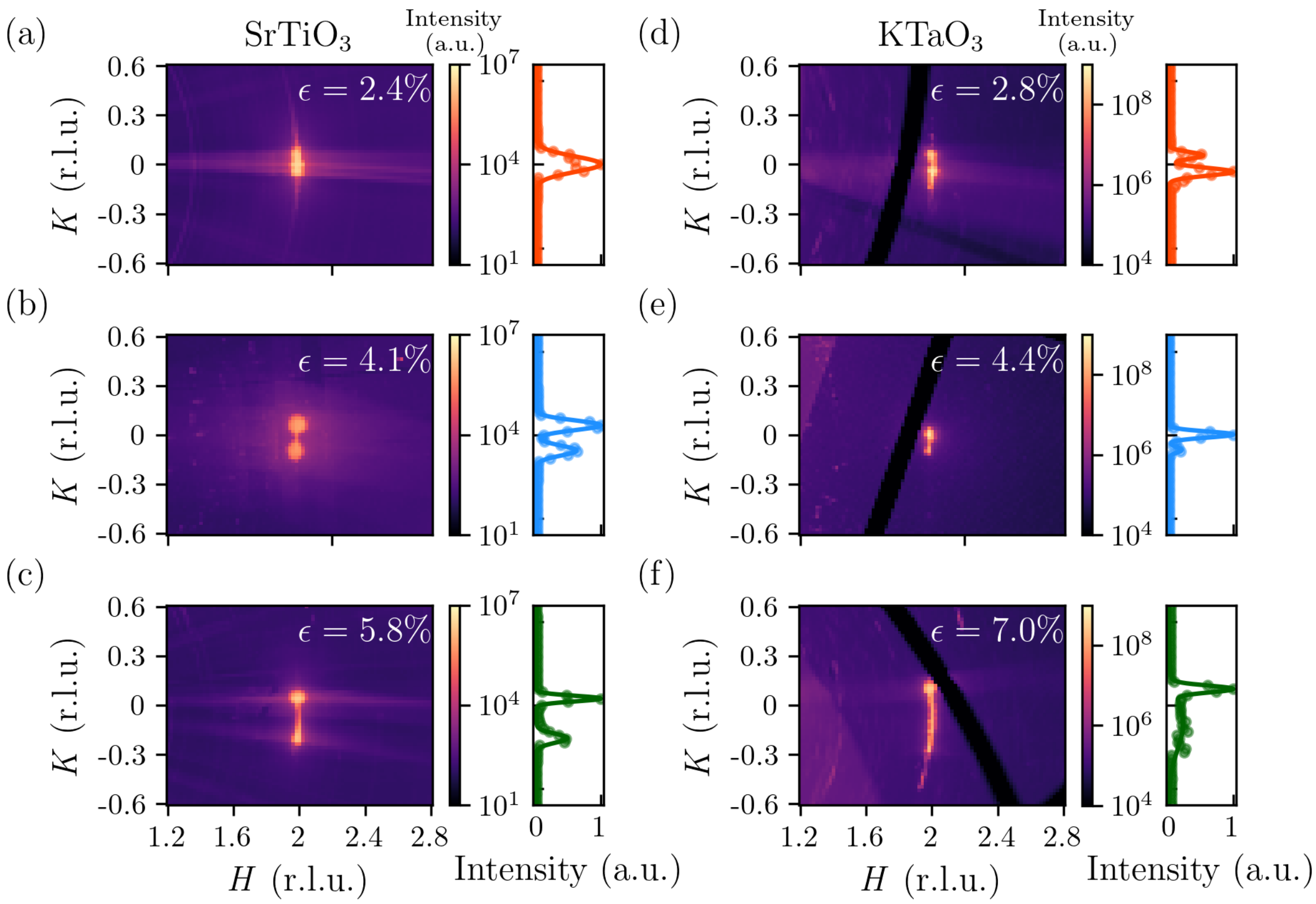

Fig. 3: Diffuse x-ray scattering near (2 0 0) for (a-c) STO (samples OVD-STO_B, OVD-STO_C, OVD-STO_D) and (d-f) KTO (samples KTO_A, KTO_B, and KTO_C) after *ex situ* deformation along [0 1 0] to the indicated strain level. To the right of each contour plot is a one-dimensional cut along [2 *K* 0] that displays the splitting of the Bragg peaks as a result of the formation of two sets of domains. Solid lines indicate best fits to one or two gaussians. There is typically a difference in intensity between the two peaks, especially in the case of KTO. The radially-oriented black streaks seen in some of the contour plots are detector artifacts.

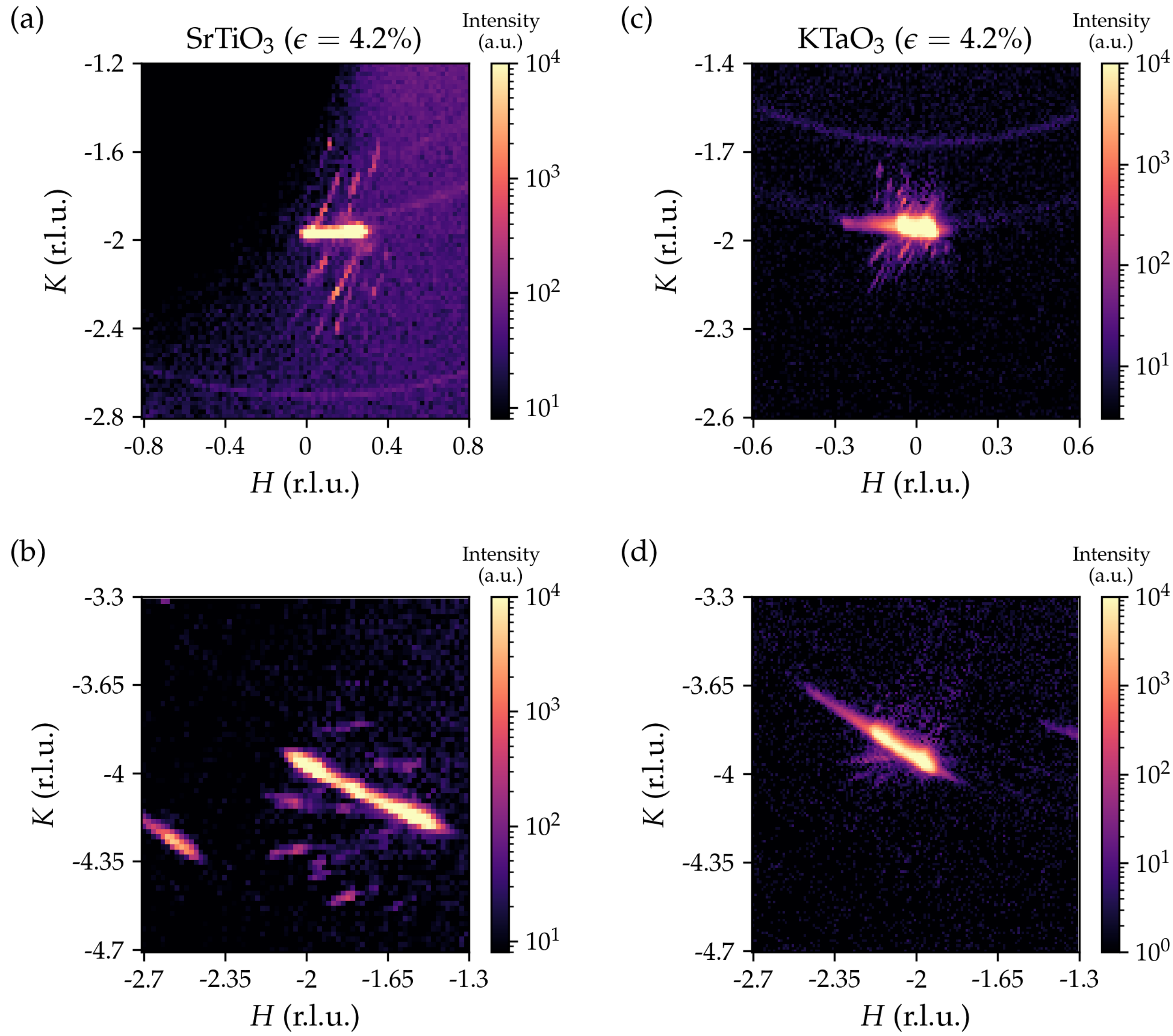

Fig. 4: Diffuse quasistatic neutron scattering for (a-b) OVD STO (sample OVD-STO_E) and (c-d) KTO (sample KTO_E) after *ex situ* deformation by 4.2% along [0 1 0]. As seen at the (0 -2 0) reflection (panels (a) and (c)) and the (-2 -4 0) reflection (panels (b) and (d)), both materials display similar cross-like patterns corresponding to an activation of the [1 -1 0] and [1 1 0] slip planes. The difference in periodicity of these patterns indicates that the Burgers vector magnitude in KTO is larger than in STO (see text).

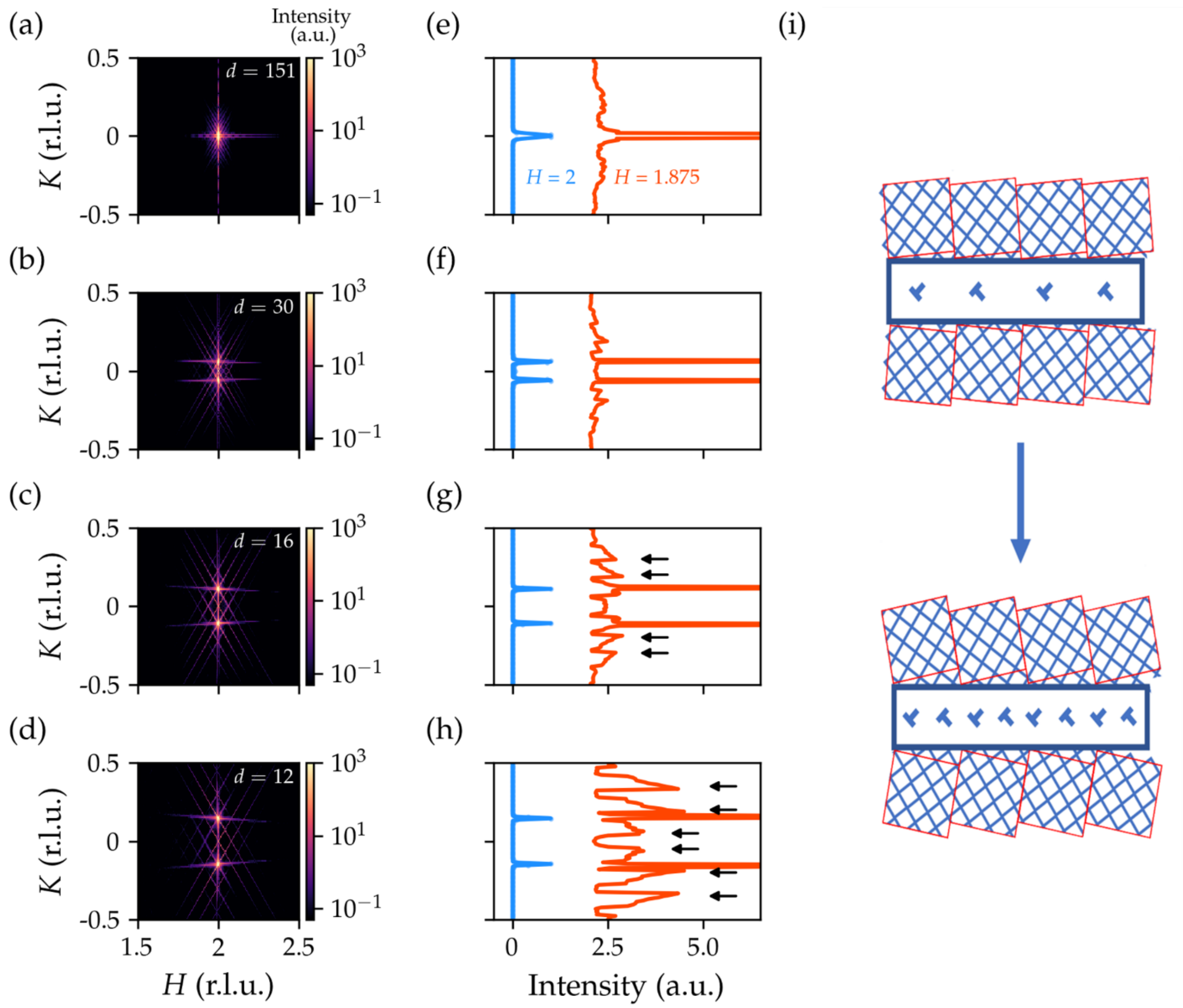

Fig. 5: Calculated two-dimensional diffuse scattering from a dislocation wall with a Burgers vector magnitude of $b = 2\sqrt{2}a$ ($a = 3.905$ Å is the room temperature lattice constant of STO) and distances between dislocations of (a) 151, (b) 30, (c) 16 and (d) 12 unit cells. The angular spread is seen to widen as the dislocation density increases. (e-h) Line cuts along [$H$ $K$ 0] for $H$ = 2 r.l.u. (blue curves) and $H$ = 1.875 r.l.u. (orange curves). The line cuts at $H$ = 2 r.l.u. are fit to gaussians to extract the parameters for the asterisms, which are included in Table S1. The line cuts at $H$ = 1.875 r.l.u., whose intensities are scaled up for clarity, display the change in spacing between diffuse streaks, as indicated by the arrows in panels (g) and (h). (i) Real-space schematic of the relation among tilted domains (blue lines represent the {1 1 0} planes), dislocation density, and strain. Two types of domains separated by dislocation walls form as the system is strained. The number of dislocations (blue, T-shaped symbols, pointing along [1 1 0]/[1 -1 0]) increases within the dislocation wall as strain increases, resulting in a larger angular tilt between the two sets of domains.

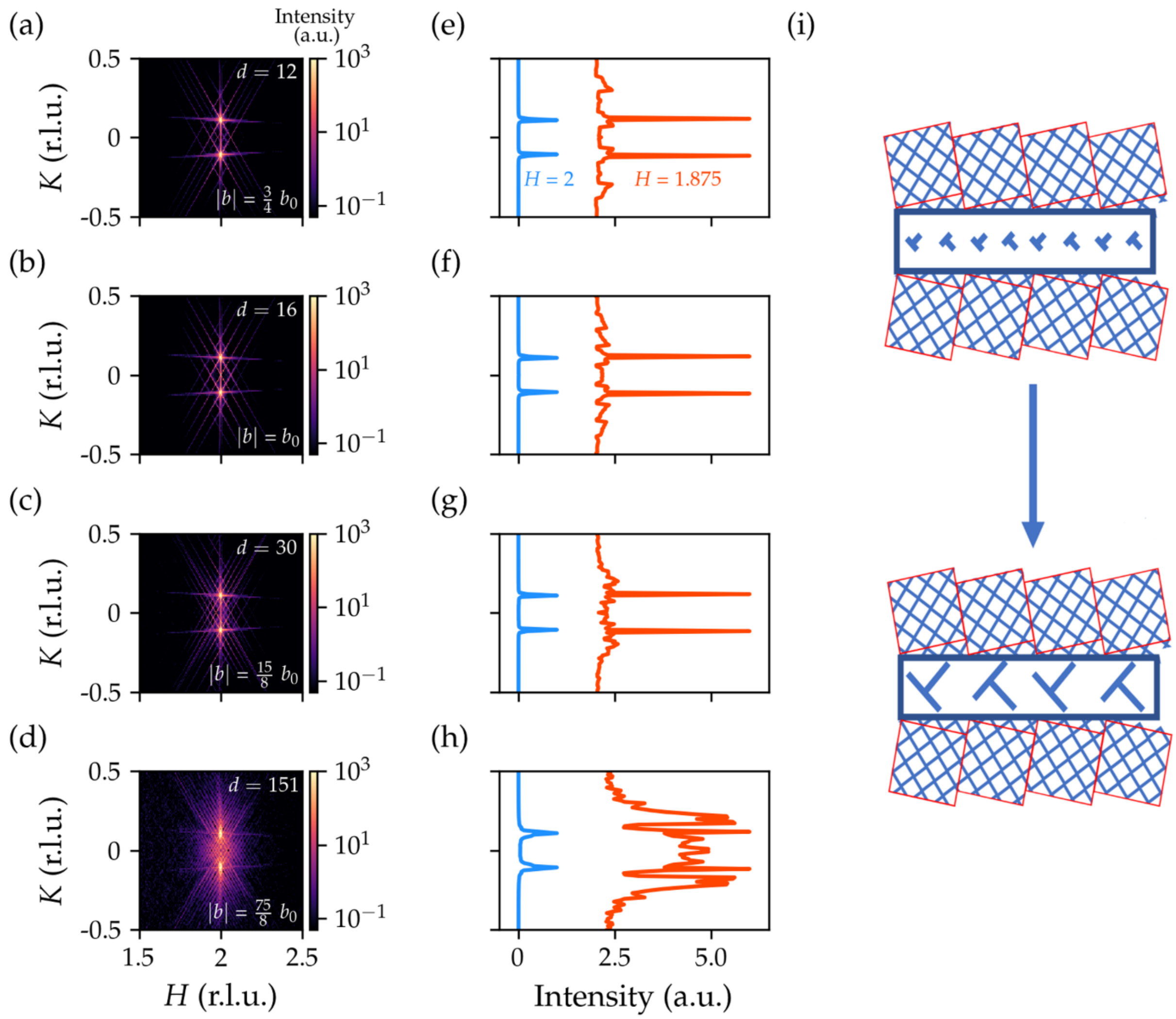

Fig. 6: Calculated diffuse scattering from a dislocation wall with distance between dislocations $d$ of (a) 12, (b) 16, (c) 30 and (d) 151 unit cells. The Burgers vector magnitude $b$ is adjusted as indicated to maintain a constant linear dislocation density, where $b_0$ has a length of $2\sqrt{2}a$. (e-h) Corresponding line cuts along $[H\ K\ 0]$ for $H$ = 2.0 and 1.875 r.l.u. Although the angular spread and domain tilt angle are unchanged, the reciprocal space periodicity of the fine structure increases with increasing Burgers vector magnitude, as indicated schematically in (e) (larger dislocations correspond to a larger Burgers vector).

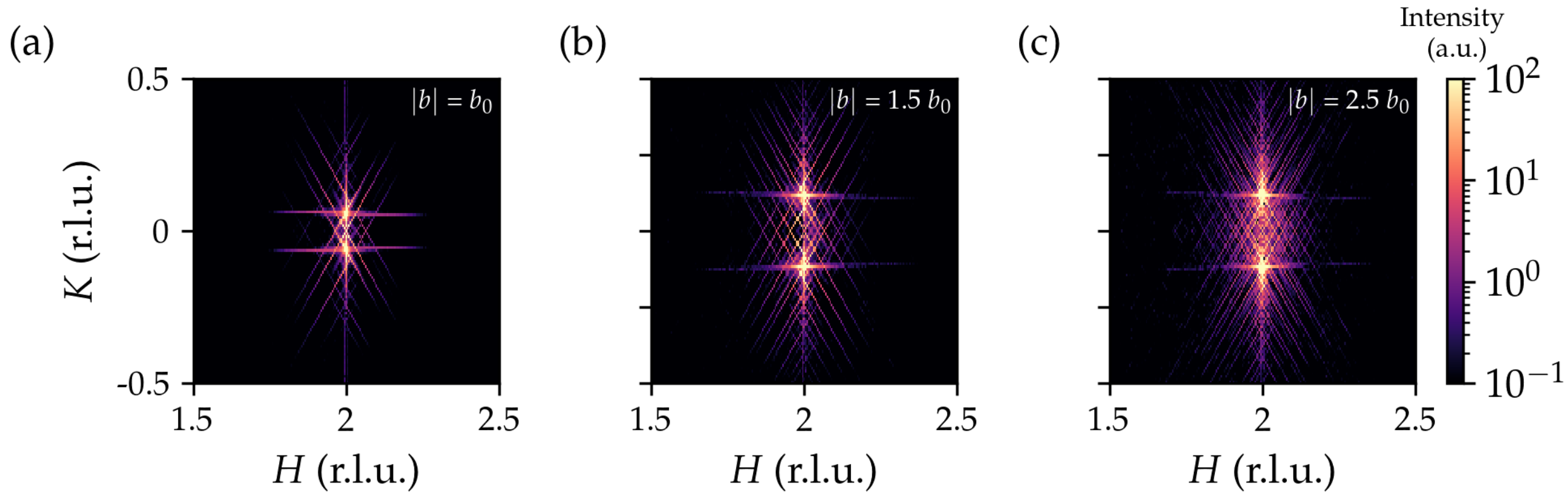

Fig. 7: (a-c) Calculated diffuse scattering from a dislocation wall of fixed length with varying Burgers vector magnitude (as indicated), used to determine which periodicity matches best with the diffuse neutron scattering pattern for 4.2% deformed KTO. The dislocation distance $d$ = 30 was found to best match the Bragg reflection splitting seen in Fig. 4 (c), and the periodicity of the streaks in the fine structure was found to be well described by $b$ = 2.5 $b_0$ (see panel (c); note that $b_0 = 2\sqrt{2}\,a$. On the other hand, the periodicity in (a) matches that observed for 4.2% deformed STO (Fig. 4 (a)), which suggests that the Burgers vector magnitude for KTO is 2.5 times larger than for STO.

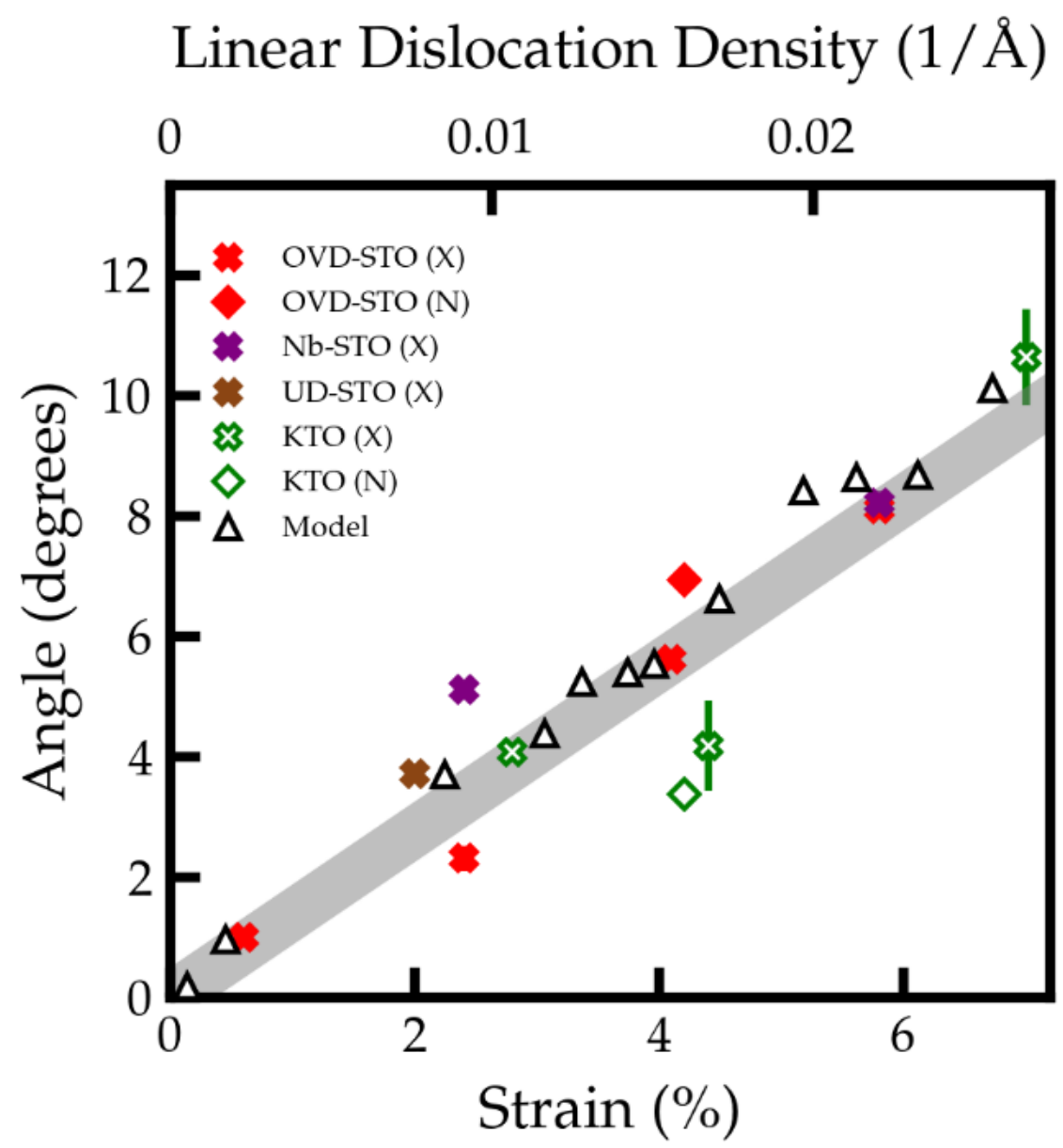

Fig. 8: Angular spread observed *via* x-ray (X) and neutron (N) diffuse scattering for deformed STO (undoped; oxygen-vacancy-doped, OVD-STO; Nb-doped, Nb-STO) and KTO (undoped) as a function of strain (bottom axis). In each case, data for more than ten Brillouin zones were analyzed (generally with $Q$ > 3 r.l.u.) *via* simple gaussian fits, as described in the text. The grey band indicates the result of a fit of the combined result to the power-law form $W = A * x^p$, where $x$ is the strain level. We find that $p$ = 1.0 $\pm$ 0.2. Additionally, the angular spread of the (2 0 0) Bragg reflection from the model calculation (black triangles) is shown as a function of linear dislocation density (top axis). The linear dislocation density values are determined as 1/($d * a$) $d$. A linear relationship is observed between the angular spread of the asterism and the linear dislocation density, indicating that strain and linear dislocation density are also linearly related.

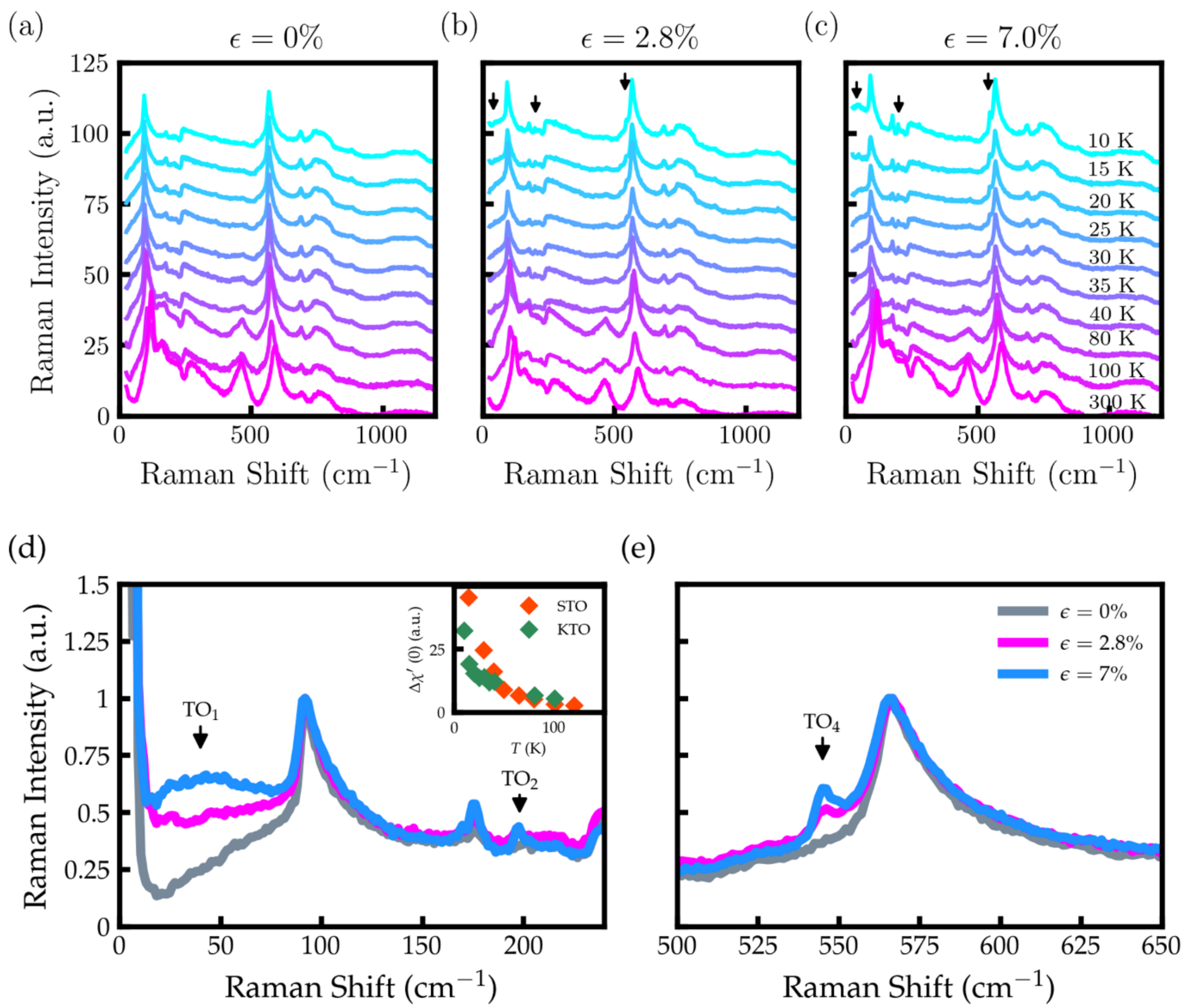

Fig. 9: Polarized Raman scattering results for (a) undeformed (KTO_D), (b) 2.8% deformed (KTO_A), and (c) 7% deformed (KTO_C) KTO, measured from 10 K to 300 K (temperatures indicated in (c)). Data were collected in *y*-*y* polarization geometry, with the polarized light at an angle of 45° relative to the [0 1 0] deformation direction. The arrows in (b) and (c) indicate three new modes that emerge after plastic deformation. (d) After deformation, a broad, low-energy feature is observed in addition to the emergence of the $TO_2$ mode at 45 cm$^{-1}$. The spectra are normalized to the peak at 100 cm$^{-1}$. The inset shows the temperature dependence of the low energy mode in KTO_C (panel (c)) and undoped STO [34]. (e) Similarly, the $TO_4$ mode just below 550 cm$^{-1}$ is seen to emerge after deformation. The spectra are normalized to the peak at 570 cm$^{-1}$. The data in (d) and (e) were taken at 10 K. The emergence of these three features signifies the presence of inversion-symmetry breaking (the $TO_2$ and $TO_4$ modes) and the presence of polar fluctuations (the broad low-energy feature at 45 cm$^{-1}$) in a non-zero volume fraction of the deformed crystals. The $1/T$ Curie dependence of the low-energy mode indicates the presence of quantum-critical fluctuations. These features appear to be enhanced at higher deformation levels. The two peaks at 100 and 570 cm$^{-1}$ are used to normalize the data and likely associated with TA phonon modes that are unaffected by high levels of strain [51].

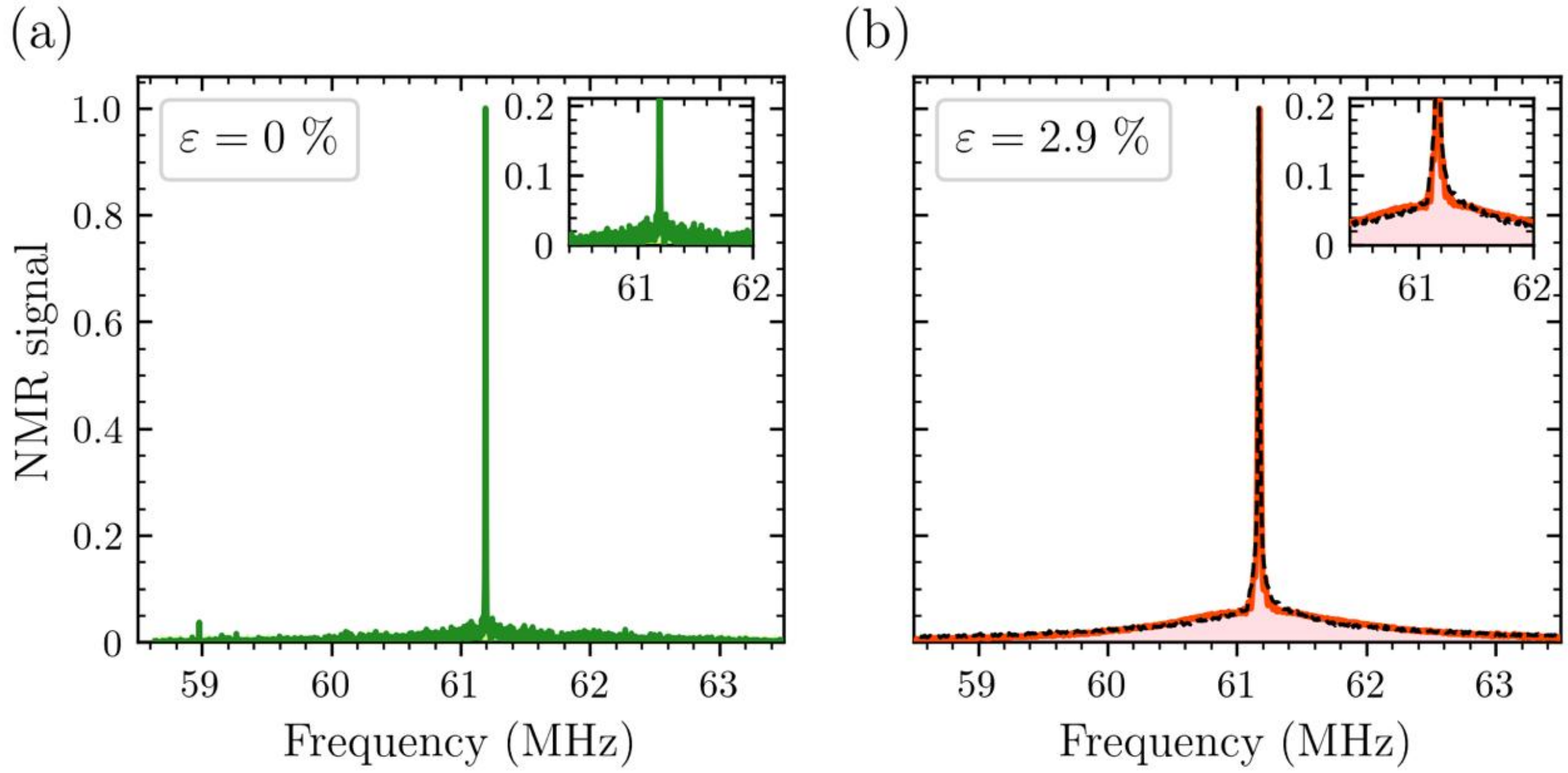

Fig. 10: [181]Ta NMR spectra in KTO (sample KTO_F) at 30 K (a) before and (b) after deformation to 2.9%. In the latter case, the spectrum exhibits two components: a narrow central line and a broad signal due to the superposition of satellite lines of nuclei with different quadrupole frequencies $\nu_Q$. Based on the integrated intensities, about 1.2% of the volume fraction is highly strained, while the majority of the sample is unaffected by the dislocation structures. The dashed black line in (b) is the result of a simulation based on the strain field around a dislocation wall (see text). The broad component is absent in the case of the undeformed sample.